\documentclass[11pt,a4paper]{article}

\usepackage[utf8]{inputenc}
\usepackage[T1]{fontenc}
\usepackage[british]{babel}
\usepackage{lmodern}
\usepackage{microtype}
\usepackage[top=2.2cm, bottom=2.2cm, left=2.4cm, right=2.4cm]{geometry}
\usepackage[dvipsnames]{xcolor}
\usepackage{booktabs}
\usepackage{tabularx}
\usepackage{colortbl}
\usepackage{tikz}
\usetikzlibrary{arrows.meta, positioning, calc, fit, backgrounds, shapes.multipart}
\usepackage{amssymb}
\usepackage{graphicx}
\usepackage{caption}
\usepackage[hidelinks]{hyperref}

\definecolor{arblue}{HTML}{3730A3}
\definecolor{arlight}{HTML}{EDE9FE}
\definecolor{stubred}{HTML}{DC2626}
\definecolor{accgreen}{HTML}{059669}
\definecolor{ingray}{HTML}{6B7280}

\tikzset{
  inchip/.style={rounded corners=2.5pt, draw=ingray!60, fill=ingray!10,
                 minimum height=0.7cm, text width=3.2cm, align=left,
                 font=\scriptsize\ttfamily, inner sep=4pt},
  outchip/.style={rounded corners=2.5pt, draw=accgreen!60, fill=accgreen!10,
                  minimum height=0.7cm, text width=3.2cm, align=left,
                  font=\scriptsize\ttfamily, inner sep=4pt},
  pbox/.style={rounded corners=3pt, draw=arblue!60, fill=white, line width=0.5pt,
               minimum height=0.95cm, minimum width=3.6cm, text width=3.4cm,
               align=center, font=\scriptsize, inner sep=3pt},
  inflow/.style={-Stealth, thick, ingray!70, shorten >=1pt},
  outflow/.style={-Stealth, thick, accgreen!70, shorten <=1pt},
  pboxflow/.style={-Stealth, semithick, arblue!70},
  skillbox/.style={rounded corners=10pt, draw=arblue, line width=1.2pt, fill=arblue!4},
  skilltitle/.style={font=\large\bfseries\sffamily, text=arblue, anchor=north west},
  level/.style={rounded corners=4pt, draw=arblue!60, fill=arblue!6, line width=0.6pt,
                minimum height=1.6cm, text width=3.6cm, align=center, font=\small, inner sep=5pt},
  levelflow/.style={-Stealth, semithick, arblue!70},
  hooknode/.style={rounded corners=3pt, draw=stubred!70, fill=stubred!8, line width=0.7pt,
                   minimum height=0.85cm, text width=2.4cm, align=center, font=\scriptsize},
  modelnode/.style={rounded corners=3pt, draw=arblue!70, fill=arblue!6, line width=0.7pt,
                    minimum height=0.85cm, text width=2.6cm, align=center, font=\scriptsize},
  toolnode/.style={rounded corners=3pt, draw=ingray!70, fill=ingray!10, line width=0.7pt,
                   minimum height=0.85cm, text width=2.2cm, align=center, font=\scriptsize},
  loopflow/.style={-Stealth, semithick, arblue!70},
  axisline/.style={-Stealth, thick, ingray!80},
  dot/.style={circle, fill=arblue, inner sep=2pt},
  dotlab/.style={font=\scriptsize, text=black, align=center},
  umlclass/.style={rectangle split, rectangle split parts=3, draw=arblue, line width=1pt,
                   fill=arblue!4, rounded corners=2pt, text width=5.6cm, align=left,
                   font=\small, inner sep=5pt},
  umlres/.style={rectangle split, rectangle split parts=2, draw=ingray!70, line width=0.7pt,
                 fill=ingray!8, rounded corners=2pt, text width=3.6cm, align=left,
                 font=\scriptsize, inner sep=4pt},
  compo/.style={{Diamond[fill=arblue]}-, semithick, arblue!80},
  dep/.style={-{Stealth}, dashed, semithick, ingray!85}
}

\title{\textbf{Authoring Agent Skills: A Software-Engineering Approach}}
\author{Giuseppe Destefanis\\[2pt]
\normalsize Department of Computer Science, University College London\\
\normalsize \texttt{g.destefanis@ucl.ac.uk}}
\date{}

\begin{document}
\maketitle

\begin{abstract}
\noindent
Agent Skills are an emerging way to extend large language model agents with reusable procedural knowledge that the agent loads on demand. Anthropic introduced Agent Skills and published the format as an open specification supported across several agent tools. This note argues that a skill is a software artefact and that its construction should follow software-engineering principles, with qualifications: single responsibility, separation of interface from implementation, low coupling, and economy in a shared token budget, together with behavioural evaluation in place of deterministic testing. Using Claude Code as the reference implementation, it describes how a skill is structured, how its contents are loaded in stages, and how to write the description on which selection depends. It places skills against the other mechanisms a developer can use to shape agent behaviour, like project memory files, slash commands, subagents, external tool connections, and hooks, and gives a rule for choosing between them based on who decides that a mechanism runs and what guarantee it provides. It then sets out an evaluation-driven authoring process, a set of patterns and faults commonly encountered in authoring, and the trust question raised by using skills from third parties. We illustrate the comparison drawn in UML class style, the loading model, the anatomy of a skill, the relative position of each mechanism, and the points at which skills and hooks act during a session.
\end{abstract}

\section*{Note on sources}

This note draws on Anthropic's official documentation\footnote{\url{https://www.anthropic.com/}} for Agent Skills, Claude Code skills, hooks, and skill authoring best practices, together with the open Agent Skills specification.

\section{Introduction}

A skill packages a workflow, a set of conventions, domain facts, or a sequence of steps that an agent should follow for a particular kind of task, so that the agent behaves as a specialist without the developer repeating the same guidance in every session. A characteristic property of a skill is that the agent can decide when to apply it, by matching the description, and that decision is probabilistic. There is no compiler or type system confirming that the right skill fired. Authoring choices therefore determine whether a skill is selected at all and whether its instructions are followed once loaded.

Anthropic introduced Agent Skills and published them as a portable format. The directory structure, the YAML frontmatter, and the staged loading model are an open specification supported across a range of agent tools~\cite{agentskills}. A skill written against the core format can usually be read by another tool that supports the specification, though tool-specific fields and behaviours may not transfer. This note uses Claude Code as the reference implementation, because it exposes the surrounding mechanisms in full, but the structure of a skill and the authoring principles apply wherever the format is supported.

\textbf{A skill is a software artefact}, and the argument of this note rests on that claim, so it is worth stating the grounds. A skill is made of ordinary files: a \texttt{SKILL.md} file of instructions, plus any scripts and reference documents it bundles. These files can be kept under version control, like the rest of a project's code. It has an interface, the description, and an implementation, the body and any bundled scripts. It is composed with other units: it bundles resources, calls tools, and can invoke other skills. It is read and executed by a machine, it is maintained as the surrounding system changes, and it can fail, by not being selected or by being followed incorrectly. These are the properties by which software is identified, so the methods used to build software apply to skills, and this note treats authoring as a design activity. Section~\ref{sec:swe} develops the consequences.

A developer assembling a real project faces a further problem that the per-skill documentation does not address directly. Several mechanisms shape agent behaviour: project memory files, slash commands, subagents, external tool connections, and hooks. They look similar on the surface and differ in what they guarantee. Choosing the wrong one is a common source of unreliable behaviour, because a requirement that must hold every time is written as advisory prose in a place that only sometimes takes effect.

This note has four aims. The first is to argue that software-engineering principles apply to skills, and to set out the comparison drawn in UML class style and its limits. The second is to set out the structure of a skill, its staged loading model, and the way a description governs selection. The third is to position skills against the other mechanisms on the same surface and give a rule for choosing between them. The fourth is to describe an authoring process based on evaluation and iteration, together with common patterns and faults. The examples use Claude Code, and the diagrams use a generic skill that does not describe any specific deployment.

\section{Skills as software artefacts}
\label{sec:swe}

The introduction set out the grounds for treating a skill as a software artefact: ordinary files that can be versioned, an interface and an implementation, composition with other units, and the ways it can fail. Many of the disciplines that apply to ordinary software therefore apply to skills, some directly and some in a modified form, and this note treats authoring as a design activity.

\begin{figure}[ht]
\centering
\resizebox{0.86\textwidth}{!}{%
\begin{tikzpicture}
\node[umlclass] (sk) at (0,0) {%
  \hfil\textbf{\guillemotleft skill\guillemotright}\hfil\\[1pt]\hfil{\bfseries\sffamily\color{arblue} release-notes}\hfil
  \nodepart{two}\raggedright \textit{interface (description)}\\[1pt] what it does and when to use it:\\ ``drafts release notes from merged pull requests; use when cutting a release or updating a changelog''
  \nodepart{three}\raggedright \textit{implementation}\\[1pt] \texttt{SKILL.md} body: ordered steps\\ bundled: \texttt{changelog-style.md}, \texttt{gather-prs.py}};

\node[umlres] (res) at (8.6,1.0) {%
  \textbf{bundled resources}
  \nodepart{two}\texttt{changelog-style.md}\\ \texttt{gather-prs.py}};

\node[umlres] (tool) at (8.6,-1.7) {%
  \textbf{external tool / other skill}
  \nodepart{two}a code host\\ (via MCP), or a called skill};

\draw[compo] (sk.east) -- node[above, font=\scriptsize, text=arblue] {composition} (res.west);
\draw[dep] (sk.east) -- node[below, font=\scriptsize, text=ingray] {\guillemotleft uses\guillemotright} (tool.west);
\end{tikzpicture}}
\caption{A skill drawn in UML class style. The name, interface, and implementation map onto the skill name, the description, and the body with its bundled files. The notation captures structure only: a skill is module-like with a static form, and the model selects it by matching its description, where a method call is dispatched deterministically on its signature.}
\label{fig:uml}
\end{figure}
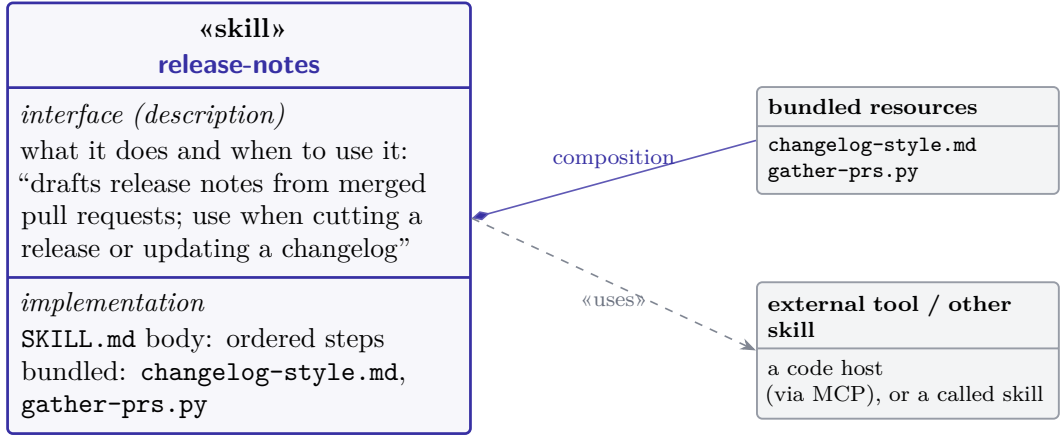

Several principles transfer, though each for a reason specific to how skills work rather than by analogy alone. A skill should have a single responsibility, one coherent capability, because the description that advertises it is matched against a task: a skill scoped to one class of task is selected more reliably, while one that does several things has a diffuse description that matches less well. The separation of interface from implementation is built into the format. The metadata that advertises a skill, its name and description, is the only part the selector reads, and the body is loaded only after selection, so the separation of interface and implementation is one of timing as well as principle. Cohesion applies in its usual sense: a skill, and each reference file within it, groups content that serves a single concern, which keeps each unit focused and, for a skill, sharpens the description that selects it. Coupling applies between skills: a skill that relies on the internal behaviour or output format of another skill, or on a shared resource, is coupled to it, so that a change to one can break the other. Such dependencies are kept few and made explicit, shown as composition and a use relation in Figure~\ref{fig:uml}, which is low coupling in the usual sense of limiting how far a change propagates. The context window is a shared resource with a fixed budget, so a skill is written to spend tokens only where they earn their place, which is resource economy under a hard limit. The name is part of the metadata the selector matches, so a consistent, descriptive name aids selection as well as readability.

Testing transfers only in a qualified form, and it is worth being exact about why. A skill cannot be unit-tested the way a function can, because it has no behaviour in isolation: it is inert text until a model reads it, so what is exercised is the model together with the skill on a task, which is integration rather than unit testing. The result is non-deterministic, so a skill is judged by a pass rate over several runs rather than a single pass or fail, and most outputs have no exact expected value, so the check is a rubric or a judge rather than an equality assertion. This is behavioural evaluation, closer to evaluating a machine-learning component than to deterministic testing, and Section~\ref{sec:eval} sets out the process. The exception is a bundled script, which is ordinary code and is unit-tested in the ordinary way.

Figure~\ref{fig:uml} draws a skill in UML class style. The top compartment holds the skill name. The interface compartment is the description, the contract a caller reads to decide whether to use the skill. The implementation compartment is the body together with the bundled files. Composition links the skill to the resources it bundles, and a dependency links it to the external tools or other skills it calls.

The comparison is a guide to structure. A skill is module-like in behaviour: a single static unit, like a class that holds only static members, and each skill is independent of the others. Its invocation is probabilistic: the model selects a skill at runtime by matching its description, while a method call is dispatched deterministically on its signature. The static form and the description-based selection are what an author should keep in mind when reading the diagram.

That last difference is the strongest reason to take the design seriously. In conventional code a type checker and a linker catch a mis-wired call before it runs. A skill has no such guarantee: a poorly named or poorly scoped skill fails quietly, either by not being selected or by being selected and then ignored. The contract and the evaluations, that is the description and the behavioural tests, carry the weight that a type system would otherwise carry. Clear interfaces, low coupling, and executable evaluations matter more here than in code a compiler can check.

Where a skill governs a task, the agent's behaviour on that task turns substantially on the quality of the skill. A skill that is selected reliably and followed faithfully raises the floor on the agent's behaviour for that task; one that is not lowers it, often without a visible error. Skill quality is therefore a determinant of system behaviour, which is the reason to architect skills with the care given to the modules of a conventional system. The rest of this note sets out that structure in detail.

\section{The skill as a unit}

A skill is a directory on the filesystem that contains a single \texttt{SKILL.md} file and, optionally, supporting files such as reference documents, scripts, and templates. The \texttt{SKILL.md} file has two parts: YAML frontmatter holding a \texttt{name} and a \texttt{description}, and a markdown body holding the instructions. A minimal skill is one \texttt{SKILL.md} with no other files. A larger skill bundles reference material and code alongside it~\cite{skills-overview}. In Claude Code, the reference implementation used throughout this note, skills live in \texttt{\textasciitilde/.claude/skills/} for personal use, or in \texttt{.claude/skills/} inside a repository for project-scoped use that travels with the codebase~\cite{skills-claudecode}.

The frontmatter has two required fields with fixed constraints. The \texttt{name} may be at most 64 characters, may contain only lowercase letters, numbers, and hyphens, may not contain XML tags, and may not contain the reserved words \texttt{anthropic} or \texttt{claude}. The \texttt{description} must be non-empty, at most 1024 characters, and free of XML tags~\cite{skills-overview}.

\section{Staged loading}

\begin{figure}[ht]
\centering
\resizebox{\textwidth}{!}{%
\begin{tikzpicture}
\node[level] (l1) at (0,0) {\textbf{Level 1}\\Metadata\\\texttt{name} + \texttt{description}};
\node[level] (l2) at (7.0,0) {\textbf{Level 2}\\\texttt{SKILL.md} body\\instructions};
\node[level] (l3) at (14.0,0) {\textbf{Level 3}\\Reference files\\and scripts};

\draw[levelflow] (l1.east) -- node[above, font=\scriptsize, text=arblue, align=center] {reads when\\relevant} (l2.west);
\draw[levelflow] (l2.east) -- node[above, font=\scriptsize, text=arblue, align=center] {reads or runs\\as needed} (l3.west);

\node[font=\scriptsize\itshape, text=ingray, align=center] at (0,-1.7) {always loaded\\$\approx$100 tokens / skill};
\node[font=\scriptsize\itshape, text=ingray, align=center] at (7.0,-1.7) {loaded on trigger\\under $\approx$5k tokens};
\node[font=\scriptsize\itshape, text=ingray, align=center] at (14.0,-1.7) {loaded on demand\\no fixed limit};
\end{tikzpicture}}
\caption{Staged loading. Only the metadata is resident at all times. The body enters context when the model selects the skill, and bundled files are read or executed only when the task requires them.}
\label{fig:loading}
\end{figure}
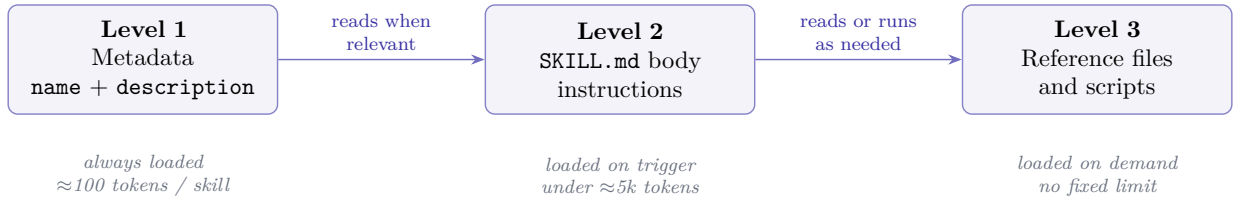

Skills use staged loading, described in the documentation as progressive disclosure~\cite{skills-overview}. The mechanism rests on three levels, shown in Figure~\ref{fig:loading}.

The first level is the frontmatter. The \texttt{name} and \texttt{description} of every installed skill are made available to the model at the start of a session, at a cost of roughly one hundred tokens per skill~\cite{skills-overview}. This is what lets a project hold many skills without paying for all of their contents up front. At this stage the model knows only that the skill exists and what it claims to be for.

The second level is the body. When the model judges that a task matches a skill's description, it reads \texttt{SKILL.md} from disk and the body enters the context window. The body should stay under about five hundred lines, a few thousand tokens, so that once loaded it does not crowd out the conversation and other context~\cite{skills-bestpractices}.

The third level is the bundled files. Reference documents are read only when the body points to them and the task needs them. Scripts are normally run rather than read into context: executing a script keeps its source out of the context window and brings in only its output. A script can instead be read as reference where its internal logic matters, in which case it counts as a reference file. There is no token cost for bundled material until it is accessed, which means a skill can carry large reference files or datasets without penalty as long as they remain unread~\cite{skills-overview}.

A skill has two invocation paths. Under automatic invocation the model decides whether to run a skill, by matching the description, and that decision is probabilistic. A skill can also be invoked directly by name, which runs it deterministically at the user's choice. The description carries a double load on the automatic path: it is both the documentation a reader sees and the trigger condition the model matches against.

\section{Writing the description}

Because the description is the trigger, it deserves more care than any other part of the skill. Write it in the third person, since it is read by the model as part of its own context, and a first or second person voice can degrade matching. State both what the skill does and when it should be used, and include the concrete terms a relevant task would contain.

A description such as ``Handles releases'' gives the model nothing to match against. A description such as ``Drafts release notes from the pull requests merged between two version tags. Use when cutting a release, updating a changelog, or summarising what changed in a version'' names the operations and the triggers, so the model can select it against a real request~\cite{skills-bestpractices}.

\paragraph{Explicit invocation.} If a skill should run only when called by name, Claude Code allows the frontmatter to disable automatic invocation, which turns the skill into an explicit command~\cite{skills-claudecode}. Use this for actions that should never fire on the model's own judgement.

\section{Anatomy of a well-formed skill}

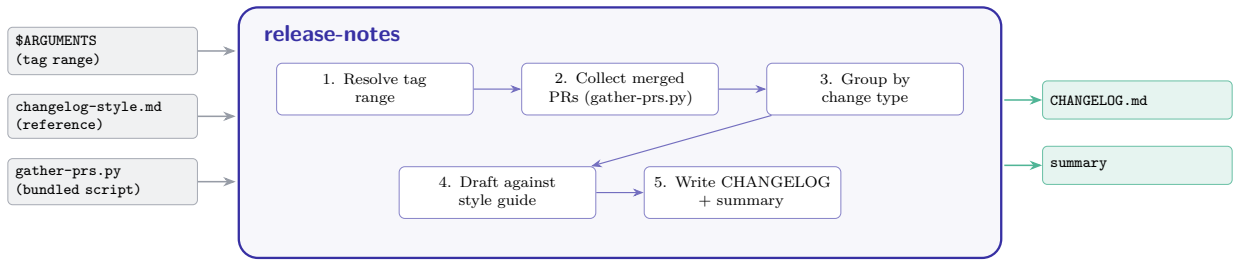
\begin{figure}[ht]
\centering
\resizebox{\textwidth}{!}{%
\begin{tikzpicture}
\node[skillbox, minimum width=14cm, minimum height=4.6cm] (sk) at (0,0) {};
\node[skilltitle] at ([xshift=10pt, yshift=-6pt]sk.north west) {release-notes};

\node[inchip] (i1) at (-9.5, 1.5)  {\$ARGUMENTS\\(tag range)};
\node[inchip] (i2) at (-9.5, 0.3)  {changelog-style.md\\(reference)};
\node[inchip] (i3) at (-9.5,-0.9)  {gather-prs.py\\(bundled script)};

\node[outchip] (o1) at (9.5, 0.6)  {CHANGELOG.md};
\node[outchip] (o2) at (9.5,-0.6)  {summary};

\node[pbox] (s1) at (-4.5, 0.8) {1. Resolve tag\\range};
\node[pbox] (s2) at ( 0,   0.8) {2. Collect merged\\PRs (gather-prs.py)};
\node[pbox] (s3) at ( 4.5, 0.8) {3. Group by\\change type};
\node[pbox] (s4) at (-2.25,-1.1) {4. Draft against\\style guide};
\node[pbox] (s5) at ( 2.25,-1.1) {5. Write CHANGELOG\\+ summary};

\draw[pboxflow] (s1) -- (s2);
\draw[pboxflow] (s2) -- (s3);
\draw[pboxflow] (s3) -- (s4);
\draw[pboxflow] (s4) -- (s5);

\foreach \i in {i1,i2,i3}
  \draw[inflow] (\i.east) -- (sk.west |- \i);
\foreach \o in {o1,o2}
  \draw[outflow] (sk.east |- \o) -- (\o.west);
\end{tikzpicture}}
\caption{Anatomy of a skill (generic example). Grey chips are inputs the skill reads, including reference files and bundled scripts; the numbered boxes are the steps encoded in the body; green chips are the artefacts produced. When a bundled script is executed rather than read, its source does not enter context.}
\label{fig:anatomy}
\end{figure}

Figure~\ref{fig:anatomy} shows the parts of a skill using a generic release-notes example. Inputs that the skill reads are drawn as chips on the left, the numbered steps of the body sit inside the container, and the artefacts the skill produces are chips on the right. The same convention serves any skill: the container is the unit of behaviour, the chips on either side are what it consumes and produces, and the inner boxes are the procedure the body encodes.

The body should be short and oriented to the task it serves. Assume the model already holds general knowledge, and add only what it does not have: project conventions, non-obvious procedures, specific file locations, rules that must hold. Every sentence competes for context once the body is loaded, so remove explanations of things the model already understands.

\paragraph{References stay one level deep.} The body may link to reference files, but those files should not chain onward to further files, because the model may preview a deeply linked file with a partial read and then act on incomplete information. Link every reference file directly from \texttt{SKILL.md}. For any reference file longer than about one hundred lines, place a short table of contents at the top so the model can see the full scope even on a partial read~\cite{skills-bestpractices}.

\paragraph{Conventions.} Use forward slashes in all paths so they work across operating systems. Use one term for each concept throughout the skill rather than alternating between synonyms, since consistent wording is easier for the model to follow. Avoid time-sensitive statements that will date; if older behaviour must be recorded, place it in a clearly marked section for legacy patterns rather than in the main flow~\cite{skills-bestpractices}.

\paragraph{Degrees of freedom.} Match the level of constraint to the fragility of the task. Where several approaches are valid and the right one depends on context, give general direction and let the model choose. Where a sequence is fragile and must run exactly, give the precise command and state that it must not be altered~\cite{skills-bestpractices}.

\section{Skills within the agent tool surface}

Skills are one of several mechanisms that change how an agent behaves. They are easy to confuse, and choosing the wrong one is a common source of unreliable behaviour. Two questions separate them: who decides that the mechanism runs, and what guarantee it provides. Table~\ref{tab:surface} sets out the mechanisms against these two questions, as Claude Code implements them. Analogous mechanisms exist in other tools: the external tool connection follows the cross-vendor Model Context Protocol (MCP), and the project memory file has a cross-tool analogue in the \texttt{AGENTS.md} convention.

\begin{table}[!ht]
\centering
\small
\renewcommand{\arraystretch}{1.25}
\begin{tabularx}{\textwidth}{l >{\raggedright\arraybackslash}X >{\raggedright\arraybackslash}X >{\raggedright\arraybackslash}X}
\toprule
\rowcolor{arlight}
\textbf{Mechanism} & \textbf{Who decides it runs} & \textbf{What it provides} & \textbf{Reach for it when} \\
\midrule
Memory file (\texttt{CLAUDE.md}) & The runtime, always & Standing project context, present every turn & The instruction is short, general, and should stay in view at all times \\
Skill & The model, by matching the description; or the user, by name & Procedural knowledge and bundled resources, loaded in stages & The task needs domain procedure or reference material the model should find on its own \\
Slash command & The user, by typing \texttt{/name} & A saved prompt run on demand & You want to decide exactly when a prompt runs \\
Subagent & The model or the user, by delegating & A separate instance with its own context window and tools & Work should run in isolation to keep the main context focused \\
External tool (MCP) & The model, by calling the tool & A connection to an external service or data source & The model must act in the outside world, on a repository, a database, or a chat system \\
Hook & The runtime, on a lifecycle event & Deterministic execution that can block an action & A step must happen every time, or an action must be prevented \\
Plugin & Installed by the user & A bundle of the above for distribution & You are packaging capabilities to share across a team \\
\bottomrule
\end{tabularx}
\caption{The mechanisms that shape agent behaviour, separated by who decides that each runs and what it guarantees.}
\label{tab:surface}
\end{table}

\paragraph{Overlap between commands and skills.} In current Claude Code, slash commands and skills have converged: both can be invoked as \texttt{/name}, and files in \texttt{.claude/commands/} continue to work~\cite{skills-claudecode}. The practical reading is that a slash command and a skill sit on one continuum. The command end is a single prompt file triggered by name. The skill end adds automatic invocation by description, staged loading, and bundled files. Choose the command end to decide when it runs, and the skill end to let the model recognise the moment or to carry supporting files and scripts.

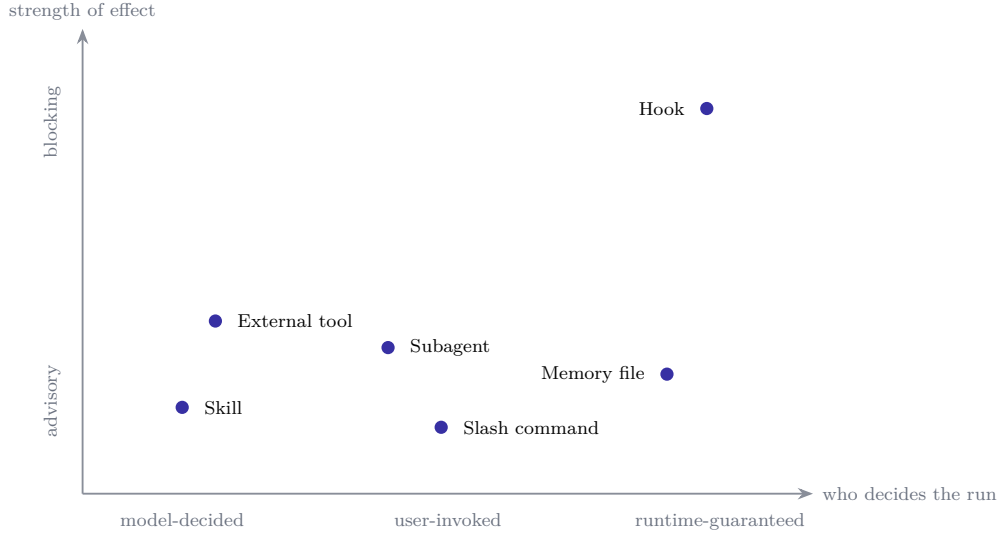
\begin{figure}[ht]
\centering
\resizebox{0.82\textwidth}{!}{%
\begin{tikzpicture}
\draw[axisline] (0,0) -- (11,0) node[right, font=\scriptsize, text=ingray] {who decides the run};
\draw[axisline] (0,0) -- (0,7) node[above, font=\scriptsize, text=ingray] {strength of effect};
\node[font=\scriptsize, text=ingray, anchor=north] at (1.5,-0.15) {model-decided};
\node[font=\scriptsize, text=ingray, anchor=north] at (5.5,-0.15) {user-invoked};
\node[font=\scriptsize, text=ingray, anchor=north] at (9.6,-0.15) {runtime-guaranteed};
\node[font=\scriptsize, text=ingray, rotate=90, anchor=south] at (-0.2,1.4) {advisory};
\node[font=\scriptsize, text=ingray, rotate=90, anchor=south] at (-0.2,5.6) {blocking};

\node[dot] (sk)  at (1.5,1.3) {}; \node[dotlab, anchor=west] at (1.7,1.3) {Skill};
\node[dot] (mcp) at (2.0,2.6) {}; \node[dotlab, anchor=west] at (2.2,2.6) {External tool};
\node[dot] (sub) at (4.6,2.2) {}; \node[dotlab, anchor=west] at (4.8,2.2) {Subagent};
\node[dot] (cmd) at (5.4,1.0) {}; \node[dotlab, anchor=west] at (5.6,1.0) {Slash command};
\node[dot] (cmd2) at (8.8,1.8) {}; \node[dotlab, anchor=east] at (8.6,1.8) {Memory file};
\node[dot] (hk)  at (9.4,5.8) {}; \node[dotlab, anchor=east] at (9.2,5.8) {Hook};
\end{tikzpicture}}
\caption{Schematic positioning of the mechanisms by who decides the run, the model, the user, or the runtime, and how strong the effect is, from advisory to blocking. The two tend to track together, since only runtime-decided mechanisms can be made deterministic: a skill sits low and left, a hook high and right. A memory file is the exception, always loaded by the runtime yet only advisory, so it sits low on the right. Among these mechanisms only a hook can block an action. Positions are indicative.}
\label{fig:positioning}
\end{figure}

Figure~\ref{fig:positioning} places the mechanisms on two axes: who decides the run, the model, the user, or the runtime, and how strong the effect is, from advisory to blocking. The two tend to track together, because only runtime-decided mechanisms can be made deterministic. A skill sits in the advisory, model-decided region, and a hook in the blocking, runtime-guaranteed region. A memory file is the exception, always loaded yet advisory. The positions are indicative rather than precise, but they show the gap that matters in practice.

\section{Skills and hooks}

This pairing deserves separate treatment, because it is where guarantees are most often misplaced. On its automatic path a skill is model-invoked and probabilistic; a hook is runtime-invoked and deterministic. A hook is a shell command, or a prompt or agent handler among others, registered against a lifecycle event in \texttt{.claude/settings.json}. It fires every time its event and matcher conditions are met, whatever the model decides. These events occur at fixed points that the runtime controls: at session start, around each tool call, and when the model finishes responding, among others. A tool call is any action the agent takes through a tool, such as reading or writing a file, running a shell command, or querying an external service. The \texttt{PreToolUse} event fires immediately before such an action runs, while the agent waits, so a hook on it can inspect the action and block it. The \texttt{PostToolUse} event fires immediately after the action completes, so a hook on it can read the result. A \texttt{PreToolUse} hook blocks the pending action either by exiting with code 2 or by returning a deny decision in its output, which makes it the place to stop a dangerous command or protect a file~\cite{hooks-reference}.

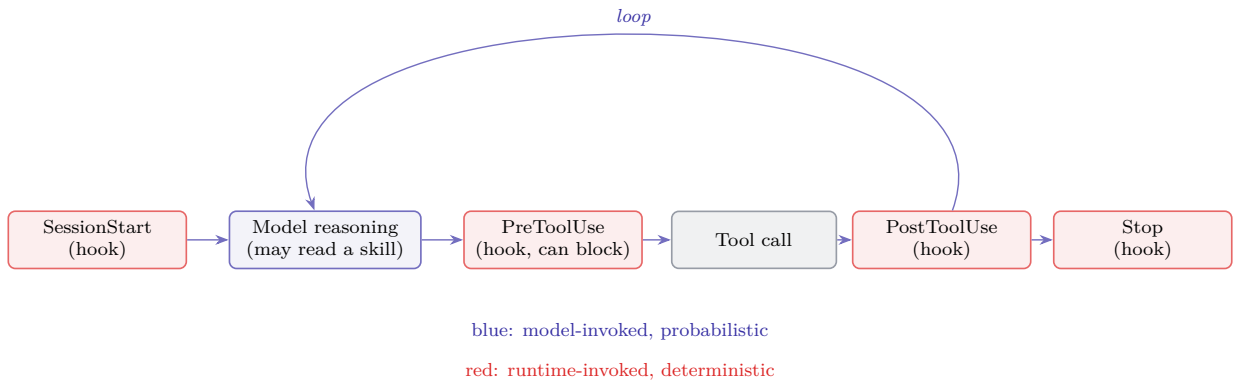
\begin{figure}[ht]
\centering
\resizebox{\textwidth}{!}{%
\begin{tikzpicture}
\node[hooknode] (start) at (0,0) {SessionStart\\(hook)};
\node[modelnode] (reason) at (3.4,0) {Model reasoning\\(may read a skill)};
\node[hooknode] (pre) at (6.8,0) {PreToolUse\\(hook, can block)};
\node[toolnode] (tool) at (9.8,0) {Tool call};
\node[hooknode] (post) at (12.6,0) {PostToolUse\\(hook)};
\node[hooknode] (stop) at (15.6,0) {Stop\\(hook)};

\draw[loopflow] (start) -- (reason);
\draw[loopflow] (reason) -- (pre);
\draw[loopflow] (pre) -- (tool);
\draw[loopflow] (tool) -- (post);
\draw[loopflow] (post) -- (stop);
\draw[loopflow] (post) to[out=70, in=110] node[above, font=\scriptsize\itshape, text=arblue] {loop} (reason);

\node[font=\scriptsize, text=arblue, anchor=north] at (7.8,-1.1) {blue: model-invoked, probabilistic};
\node[font=\scriptsize, text=stubred, anchor=north] at (7.8,-1.7) {red: runtime-invoked, deterministic};
\end{tikzpicture}}
\caption{Where skills and hooks act during a session. The model reads a skill at a point of its own choosing inside its reasoning. Hooks fire at fixed lifecycle events controlled by the runtime: \texttt{SessionStart} at the beginning of a session, \texttt{PreToolUse} and \texttt{PostToolUse} immediately before and after each tool call, with \texttt{PreToolUse} able to block the call, and \texttt{Stop} when the model finishes responding.}
\label{fig:loop}
\end{figure}

Figure~\ref{fig:loop} shows where each acts during a session. A skill is read inside the model's reasoning, at a point the model chooses. A hook fires at fixed events that the runtime controls, around every tool call and at the session and turn boundaries.

The decision rule follows directly. If a step depends on judgement, varies with context, or encodes domain procedure, put it in a skill. If a step must happen every time the triggering event occurs, put it in a hook. A standing instruction in a skill body, however firmly worded, is something the model may read, defer, or skip. The same instruction expressed as a hook runs unconditionally. When a guarantee matters, for example loading a specification file at session start, running a validator before a commit, or formatting after every file write, write it as a hook and do not rely on skill prose to enforce it.

\section{An evaluation-driven authoring process}
\label{sec:eval}

Write evaluations before writing the skill. Run the model on representative tasks without the skill and record where it fails or lacks context. Turn those failures into a small set of test cases with expected behaviours. Measure the baseline without the skill, then write the minimum instructions needed to pass the tests, and iterate against them. This keeps the skill aimed at real gaps rather than imagined ones~\cite{skills-bestpractices}.

A reliable development loop uses two instances of the model~\cite{skills-bestpractices}. Work with one instance to draft and refine the skill, drawing on the context that would otherwise be typed by hand. Test the result with a fresh instance that has only the skill loaded, and observe its behaviour on real tasks: whether it triggers when expected, follows the references, and applies the rules. Bring specific observations back to the drafting instance and revise. The frontmatter is the first thing to check when a skill fails to trigger, since the model selects on the name and description.

\paragraph{Observe how the skill is read.} If the model reads files in an order that was not expected, the structure may be less clear than assumed. If it ignores a bundled file, that file may be unnecessary or poorly signalled. If it returns to the same file repeatedly, that content may belong in the body. Test with each model intended for use, since a body that suits a stronger model may need more detail for a faster one.

\section{Patterns and faults}

For multi-step tasks, give the body an explicit ordered workflow, and for fragile sequences include a checklist the model can copy and tick off as it proceeds. Where output quality depends on validation, build a loop: run a check, fix what it reports, run it again, and proceed only when it passes. For batch or destructive operations, have the model write a plan to a structured file and validate that file with a script before any change is applied, so that errors are caught before they take effect. Where output format matters, provide a template; where style matters, provide worked input and output examples rather than describing the style in the abstract.

Several patterns commonly cause trouble, drawn from the cited best practices and from common experience. Offering many alternatives leaves the model to choose without grounds, so give one default and name the exception. Windows-style backslash paths break on other systems. Deeply nested references lead to partial reads. Assuming a package is installed fails when it is not, so state dependencies explicitly. In bundled scripts, handle error conditions rather than failing and leaving the model to recover, and justify every constant rather than leaving unexplained values.

\section{Trust and third-party skills}

A skill should be treated as a software dependency. It can carry instructions, scripts, references, and links to external content, and the agent acts on those materials with whatever permissions it holds in the current environment. A skill that fetches external content is a particular concern, since that content can itself carry instructions and can change after the skill was first trusted. So a skill from a third party should be inspected before use: read the \texttt{SKILL.md}, the bundled scripts, the referenced resources, and any external URLs, and check each against the skill's stated purpose. This matters most when the skill can read local files, call tools, or send data outside the project~\cite{skills-overview}.

\section{Summary of guidance}

A skill is a software artefact, and the guidance above is an application of ordinary engineering discipline to it. The comparison drawn in UML class style holds for structure, while invocation works differently: the model selects a skill probabilistically, so evaluations carry the weight a type system would in conventional code. A skill is selected by the model on the strength of its description and loaded in stages, so the description must state both function and trigger, and the body must stay short and assume general knowledge. Reference files are linked one level deep, and scripts are bundled for execution rather than for reading. The choice between a skill and the other mechanisms turns on two questions: who should decide that it runs, and how strong the guarantee must be. Judgement-dependent procedure belongs in a skill; a requirement that must hold every time belongs in a hook. Authoring proceeds by writing evaluations first, then the minimum instructions needed to pass them, and refining against observed behaviour. Because a skill may contain instructions and code that the agent can act on, a skill from a third party should be read in full before it is used.

\end{document}